\documentclass[12pt]{article}
\pdfoutput=1
\usepackage[english]{babel}
\usepackage[utf8]{inputenc}
\usepackage[T1]{fontenc}
\usepackage{amsmath}
\usepackage{amsthm}
\usepackage{amsfonts}
\usepackage{url}
\usepackage{authblk}
\usepackage{tensor}
\usepackage[colorlinks = true,
	    linkcolor = blue, 
	    linkcolor = blue,
            urlcolor  = blue,
            citecolor = blue,
            anchorcolor = blue]{hyperref}
\usepackage{graphicx}
\usepackage{mathrsfs}
\usepackage{enumitem}
\usepackage{subcaption}
\usepackage[numbers]{natbib}

\usepackage[usenames,dvipsnames]{xcolor}

\newcounter{mnotecount}[section]

\newcommand{\beq}{\begin{eqnarray}}
\newcommand{\eeq}{\end{eqnarray}}
\newcommand{\ben}{\begin{eqnarray*}}
\newcommand{\een}{\end{eqnarray*}}

\newtheorem*{theorem*}{Theorem}
\theoremstyle{definition}

\newcommand{\ha}{{\hat{\alpha}}}
\newcommand{\hb}{{\hat{\beta}}}

\begin{document}
\title{Example of cross-polarized standing gravitational waves}
\author[1]{Krzysztof Głód}
\author[1]{Szymon Sikora}
\author[1]{Sebastian J. Szybka}
\affil[1]{Astronomical Observatory, Jagiellonian University}
%\affil[3]{Copernicus Center for Interdisciplinary Studies}
\date{}
\maketitle{}
\begin{abstract}
We use a cosmological counterpart of the cylindrical Halilsoy solution to illustrate properties of cross-polarized standing gravitational waves.
\end{abstract}

\section{Introduction}

The linearized gravitational waves are well understood within general relativity. In the era of gravitational wave astronomy, it is especially interesting to extend our intuitions beyond the linear regime where unexplored phenomena may be hidden. One of the most basic settings are standing gravitational waves. The studies of this type have been initiated by Bondi \cite{Hermann:2004} and Stephani \cite{Stephani:2003}. More recently, the problem has been investigated in Refs.\ \cite{Szybka:2019, SzybkaNaqvi:2021}. 

The aim of this work is to clarify the role of polarization in the context of standing gravitational waves. As a toy model, we examine a class of exact solutions to Einstein equations which are cosmological counterparts of the Halilsoy cylindrical spacetimes \cite{Halilsoy:1988}. (The Halilsoy solutions are cross-polarized Einstein-Rosen waves.) These solutions correspond to three-torus Gowdy models. In Gowdy models (contrary to the original cylindrical Halilsoy spacetime), a privileged class of stationary observers exists (at antinodes) and one can examine properties of waves relative to this class of observers. 

The three-torus Gowdy models have been already investigated in the context of standing gravitational waves \cite{SzybkaNaqvi:2021}. The solution presented in \cite{SzybkaNaqvi:2021} is a special case of the class studied here. It corresponds to + polarization of the waves. The exact nonlinear gravitational waves studied in our article have, in general, a nontrivial cross term [$\omega$ in the metric \eqref{metric}]. Following Halilsoy \cite{Halilsoy:1988}, we call this type of solutions cross polarized ($\times$ polarization). However, a clarification of the terminology is needed. Traditionally, a Gowdy class is split into polarized and unpolarized models. The solutions studied here correspond to unpolarized Gowdy models and the solution studied in \cite{SzybkaNaqvi:2021} corresponds to the polarized Gowdy model. The majority of unpolarized models are studied numerically, in contrast to the polarized case for which one may write down a complete description of the spacetime. The exact solution presented here provides a notable exception to this rule and, as such, it is interesting on its own.

The possible polarization states of the standing waves are well known in electromagnetism \cite{gpw}. Electromagnetic waves could be polarized, cross polarized, or could correspond to a mixture of both polarizations (e.g.\ an appropriate combination of both polarizations could lead to circularly polarized solutions). For electromagnetic standing waves things get more complicated. A standing wave formed by a superposition of two counterpropagating collinearly polarized waves has the same linear polarization along the wave axis. If the superposed waves are linearly, but not collinearly polarized, then the standing wave is linearly polarized at nodes and antinodes (different polarizations) and circularly polarized between them with sequentially changed handedness. If the counterpropagating waves are circularly polarized with opposite handedness, then the circular polarization is maintained along the standing wave. If the superposed waves are circularly polarized with the same handedness, then the standing wave is linearly polarized at each point with the direction of polarization periodically varying along the wave axis. In Ref.\ \cite{gpw}, standing waves with a polarization angle oscillating periodically along the wave axis (the second and the latter case) is called the polarization standing wave. The analogy between electromagnetic standing waves and gravitational standing waves implies that the nontrivial cross term $\omega$ in the metric \eqref{metric} (cross-polarized solutions in the Halilsoy terminology, denoted with $\times$) may lead to a several different behaviors of polarization. The covariant classification of gravitational standing waves is beyond the scope of this article, but we present evidence that the gravitational standing waves studied in this paper are polarization gravitational standing waves.

In this paper, we study a geodesic deviation equation in a toy model with nontrivially polarized gravitational waves. However, as pointed out to us in a private communication with Halilsoy, the topic studied here is also interesting in the broader context. The tidal forces in the Earth-Moon system are low-frequency gravitational waves \cite{tidal} and their cross-polarized components may have detectable observational consequences---the secondary tides.

\section{Setting}

We consider the line element of the Gowdy form (its relation to the Halilsoy solution is described in Appendix \ref{A1}),
\begin{equation}\label{metric}
g=e^{f}\left(-dt^2+dz^2\right)+te^{p}(dx+\omega dy)^2+te^{-p}dy^2\;,
\end{equation}
where $t>0$, $0\leq x,y,z <2\pi$ and $x^\alpha=(t,z,x,y)$. The metric functions $f$, $p$ and $\omega$ depend on $t$ and $z$ only and are given by
\begin{equation}
\begin{split}\label{sol}
e^{-p}=&t\left(\cosh(\alpha)\cosh[2\beta\sqrt{\lambda}J_0(\frac{t}{\lambda})\sin(\frac{z}{\lambda})]-\sinh[2\beta\sqrt{\lambda}J_0(\frac{t}{\lambda})\sin(\frac{z}{\lambda})]\right)\;,\\
f=&-\ln{t}-p+\;\frac{\beta^2}{\lambda}t^2\left[J_0^2(\frac{t}{\lambda})+J_1^2(\frac{t}{\lambda})-2\frac{\lambda}{t}J_0(\frac{t}{\lambda})J_1(\frac{t}{\lambda})\sin^2(\frac{z}{\lambda})\right]\;,\\
\omega=&2\beta\sqrt\lambda\sinh(\alpha) t J_1(\frac{t}{\lambda})\cos(\frac{z}{\lambda})\;,
\end{split}
\end{equation}
where $\alpha$ is a constant and $J_0$ and $J_1$ are the Bessel functions of the first kind and orders $0$ and $1$, respectively. The solution \eqref{sol} satisfies the vacuum Einstein equations and corresponds to the three-torus Gowdy model.
The number of waves on the torus is given by $1/\lambda$ where $\lambda$ is a parameter such that $1/\lambda\in\mathbb{N}$. For $\alpha=0$, the solution \eqref{sol} reduces to the + polarized case studied in \cite{SzybkaNaqvi:2021}. For $\beta=0$, it corresponds to the three-torus identified Minkowski metric in coordinates expanding along $\partial_y$. Only nondiagonal components of the metric depend on the sign of $\alpha$ (the function $\omega$ is odd in $\alpha$), so without the loss of generality we assume $\alpha\geq 0$.

In this article, in addition to the coordinate basis $\partial_\alpha$, we use two sets of nonholonomic bases. They are given by
\begin{itemize}
\item the orthonormal tetrad $e_{\hat{\alpha}}$,
\begin{equation}\label{basis}
\begin{split}
e_{\hat{0}}=&e^{-f/2}\,\partial_t\;,\quad e_{\hat{1}}=e^{-f/2}\,\partial_z\;,\\
e_{\hat{2}}=&\frac{e^{-p/2}}{\sqrt{t}}\,\partial_x\;,\quad e_{\hat{3}}=\frac{e^{p/2}}{\sqrt{t}}\,\left(-\omega\,\partial_x+\partial_y \right)\;,
\end{split}
\end{equation}
\item and the null tetrad $w_{\breve{\alpha}}=\{k,l,m,\bar m\}$,
\begin{equation}
\begin{split}\label{tetrad}
k=&\frac{1}{\sqrt 2}(e_{\hat 0}+e_{\hat 1})\;,\quad l=\frac{1}{\sqrt 2}(e_{\hat 0}-e_{\hat 1})\;,\\
m=&\frac{1}{\sqrt 2}(e_{\hat 2}-i e_{\hat 3})\;,\quad \bar m=\frac{1}{\sqrt 2}(e_{\hat 2}+i e_{\hat 3})\;,
\end{split}
\end{equation}
where $k\cdot l=-1\;,\;m\cdot\bar m=1\;,\;k\cdot k=l\cdot l=m\cdot m=\bar m\cdot\bar m=0$, and $g_{\breve{\alpha}\breve{\beta}}=-2k_{(\breve{\alpha}}l_{\breve{\beta})}+2m_{(\breve{\alpha}} \bar m_{\breve{\beta})}$.
\end{itemize}

\section{Geodesic deviation}\label{sec:3}

Our research extends results obtained previously \cite{SzybkaNaqvi:2021} to the cross-polarized standing gravitational waves (the polarization $\times$). 
A particular example of such nontrivially polarized waves is provided by the metric \eqref{metric} with the auxiliary functions given by \eqref{sol} and the nonzero value of the parameter $\alpha$.
In this section, we clarify how this parameter alters the effect of standing gravitational waves on test particles. 

The behavior of test particles is investigated relative to a freely falling observer. Similar to the case studied in \cite{SzybkaNaqvi:2021}, a preferred coordinate system exists in which nodes and antinodes are labeled by one of coordinates.\footnote{The nodes and antinodes of the gravitational waves correspond to nodes and antinodes of the metric functions $p$ and $f$ which coincide with antinodes and nodes of $\omega$, respectively.} In Appendix \ref{A3}, we show that, as in the + polarized model, observers at antinodes are stationary, namely, the curves $\gamma_{k\in\mathbb{Z}}$: $z=\lambda\pi(1/2+ k)$, $x=x_0$, $y=y_0$ with constants $x_0$ and $y_0$ are future-directed timelike geodesic. The vector tangent to this geodesic corresponds to $e_{\hat 0}$. Thus $e_{\hat 0}$ is parallelly transported along $\gamma_k$: $\nabla_{e_{\hat 0}}e_{\hat 0}=0$. Moreover, we have along $\gamma_k$: $\nabla_{e_{\hat 0}}e_{\hat\alpha}=\omega^{\hat\beta}_{\;\hat\alpha}(e_{\hat 0})e_{\hat\beta}=0$, where $\omega^{\hat\beta}_{\;\hat\alpha}$ are connection 1-forms given in Appendix \ref{A5}. Therefore, the remaining orthonormal basis vectors $e_{\hat i}$ are also parallelly transported along $\gamma_k$, which implies that $e_{\hat\alpha}$ [given by \eqref{basis}] is a freely falling frame at antinodes.

Let $\xi$ be a deviation vector, then along $\gamma_k$ in a freely falling frame
\begin{equation}\label{gde}
        \frac{d^2\xi^\ha}{d\tau^2}=-R^\ha_{\;\;\hat{0}\hb\hat{0}}\xi^\hb\;,
\end{equation}
where $\tau$ is a proper time of the observer. In vacuum, the Riemann and Weyl tensors are equal, so $R^\ha_{\;\;\hat{0}\hb\hat{0}}=C^\ha_{\;\;\hat{0}\hb\hat{0}}$. The nonzero components of the Riemann tensor (Appendix \ref{A7}) may be calculated from the curvature 2-forms presented in Appendix \ref{A6} using the relation $\Omega^{\hat\alpha}_{\;\hat\beta}=\frac{1}{2}R^{\hat\alpha}_{\;\hat\beta\hat\sigma\hat\delta}\theta^{\hat\sigma}\wedge\theta^{\hat\delta}$, where the cobasis $\theta^{\hat\alpha}$ is given in Appendix \ref{A4}. However, it is instructive to rewrite the geodesic deviation equation in terms of the complex Weyl coefficients (defined in a standard way \cite{SzybkaNaqvi:2021}). (The appropriate formulas for Weyl coefficients are too large to be usefully cited here, but one may evaluate them with the help of the computer system {\sc Mathematica}.) The relevant nonzero components of the Riemann tensor are
\begin{equation}
\begin{split}\label{Rpsi}
R^{\hat{1}}_{\;\;\hat{0}\hat{1}\hat{0}}&=2 \Re(\Psi_2)\;,\\
R^{\hat{2}}_{\;\;\hat{0}\hat{2}\hat{0}}&=\frac{1}{2} \Re\left[-2\Psi_2+\Psi_0+\Psi_4\right]\;,\\
R^{\hat{3}}_{\;\;\hat{0}\hat{3}\hat{0}}&=\frac{1}{2} \Re\left[-2\Psi_2-\Psi_0-\Psi_4\right]\;,\\
R^{\hat{2}}_{\;\;\hat{0}\hat{3}\hat{0}}&=\frac{1}{2} \Im\left[\Psi_0-\Psi_4\right]\;,
\end{split}
\end{equation}
where $\Re$ and $\Im$ denote real and imaginary parts.
At antinodes $\Psi_1=\Psi_3=0$ and $\Psi_4=\Psi_0$, 
hence $R^{\hat{2}}_{\;\;\hat{0}\hat{3}\hat{0}}=R^{\hat 3}_{\;\;\hat{0}\hat{2}\hat{0}}=0$.
For $\alpha=0$, $\Psi_0$ and $\Psi_2$ are real (not only at antinodes).
The geodesic deviation equation can be written at antinodes as
\begin{equation}
\begin{split}
\frac{d^2\xi^{\hat 0}}{d\tau^2}=&0\;,\\
\frac{d^2\xi^{\hat 1}}{d\tau^2}=&-2\Re(\Psi_2)\xi^{\hat 1}\;,\\
\frac{d^2\xi^{\hat 2}}{d\tau^2}=&\Re(\Psi_2-\Psi_0)\xi^{\hat 2}\;,\\
\frac{d^2\xi^{\hat 3}}{d\tau^2}=&\Re(\Psi_2+\Psi_0)\xi^{\hat 3}\;.
\end{split}
\end{equation}
The polarization of waves is not directly observable at antinodes because the metric function $\omega$ equals zero there. The vanishing of $R^{\hat{2}}_{\;\;\hat{0}\hat{3}\hat{0}}$ and $R^{\hat 3}_{\;\;\hat{0}\hat{2}\hat{0}}$ implies that $\frac{d^2\xi^{\hat \alpha}}{d\tau^2}\sim\xi^{\hat \alpha}$. The equations decouple and the standard effect of cross polarization is not visible at antinodes. However, the parameter $\alpha$ changes the global evolution of spacetime  and in this sense, it alters the trajectories of test particles at antinodes (via $\Psi_0$ and $\Psi_2$). The Tissot diagrams are very similar to those presented in \cite{SzybkaNaqvi:2021} and we do not include them here.

\section{Geodesic null congruences}

The geodesic deviation equation studied at antinodes in the previous section does not reveal the effect of polarization on test particles directly. The orthonormal frame \eqref{basis} is not freely falling beyond antinodes which complicates the studies in the remaining regions of spacetime. In this section, we use different method to show the effect of nontrivial polarization of standing gravitational waves on test particles. We investigate the behavior of massless test particles instead of massive test particles. We show that the polarization of gravitational waves alters shear axes of geodesic null congruences parallel to the $z$ axis. 

The symmetry of the spacetime implies that the vectors $k$ and $l$ are tangent to null congruences. Since our standing wave spacetime may be seen as a ``nonlinear'' superposition of two gravitational waves moving in opposite directions, the null tetrad \eqref{tetrad} is well adapted to the problem (a symmetry between $k$ and $l$). However, it is easy to see that $k$ and $l$ do not correspond to an affine parametrization. Therefore, it is more convenient to utilize boost freedom and introduce an alternative tetrad with $k$ and $l$ substituted by $k'=e^{-f/2}k$ and $l'=e^{f/2}l$. The spin coefficients of the new Newman-Penrose tetrad that will be useful for us are
\begin{equation}\label{eq:sc}
\begin{split}
	\kappa=&m\cdot\nabla_{k'} k'=0\;,\\
	\rho=&m\cdot\nabla_{\bar m}k'=\frac{1}{2\sqrt{2} t}e^{-f}\;,\\
	\sigma=&m\cdot\nabla_m k'=\frac{1}{2\sqrt{2}}e^{-f}\left[p_{,t}+p_{,z}+i e^p(\omega_{,t}+\omega_{,z})\right]\;,\\
	\epsilon=&\frac{1}{2}(l\cdot\nabla_{k'} k'-\bar m\cdot\nabla_{k'} m)=\frac{i}{4\sqrt{2}}e^{-f+p}(\omega_{,t}+\omega_{,z})\;.\\
\end{split}
\end{equation}
The equality $\kappa=0$ implies that $k'$ is tangent to geodesic null congruences and $\Re\epsilon= 0$ implies an affine parametrization. In this setting, $-\Re\rho$, $\Im\rho$, and $|\sigma|$ correspond to the expansion, twist (vanishing for the congruence $k'$) and shear, respectively. The argument of $\sigma$ describes the shear axes. For $\alpha=0$ (the + polarized model), $\omega=0$ and the shear axes do not rotate. 

\begin{figure}[t!]
\begin{center}
\begin{subfigure}{.5\textwidth}
%\centering
\begin{flushleft}
\includegraphics[width=10cm,angle=0]{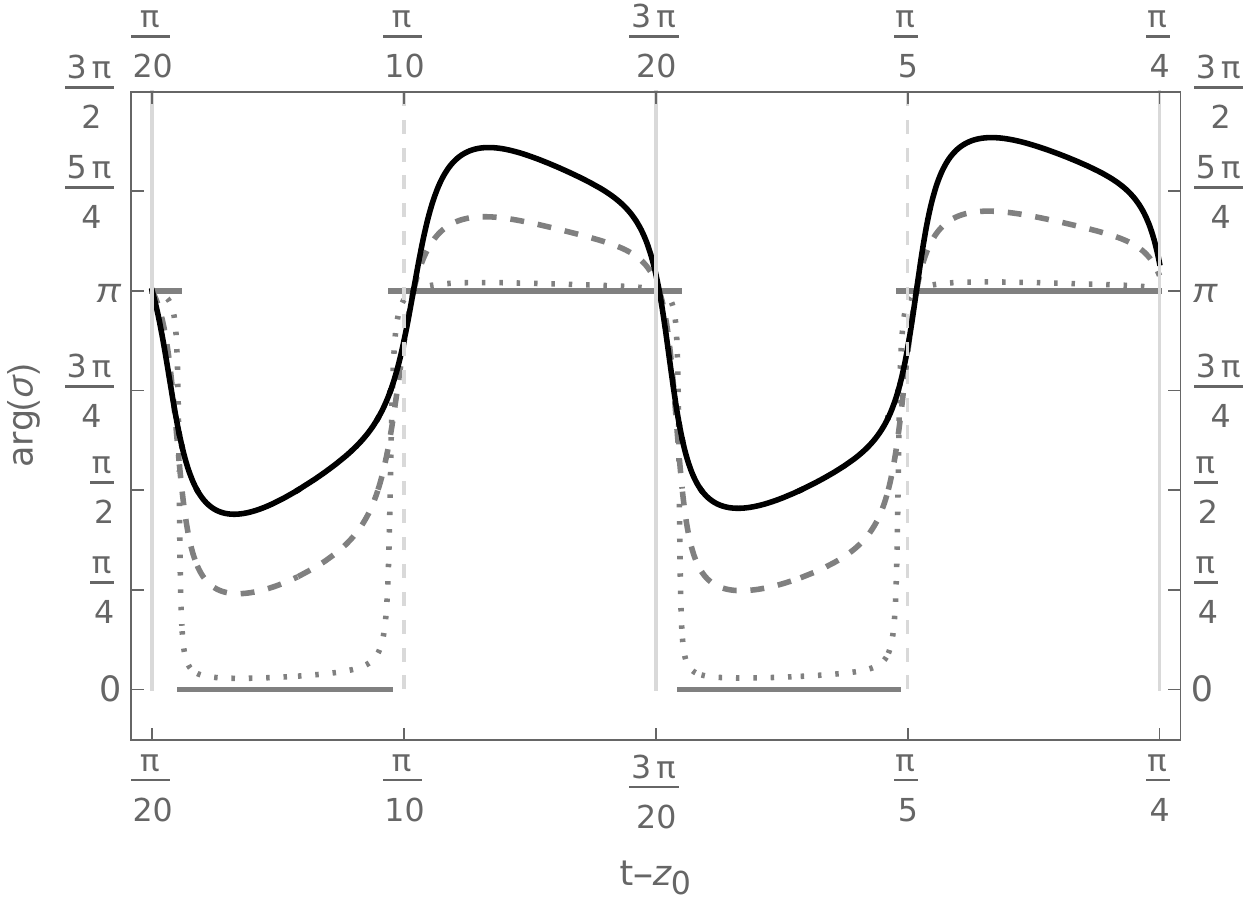}
\end{flushleft}
\end{subfigure}%
\begin{subfigure}{.5\textwidth}
\begin{flushright}
\includegraphics[width=3.3cm,angle=0]{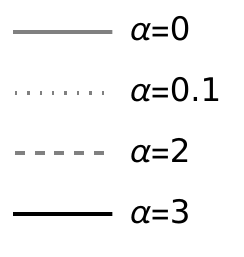}
\end{flushright}
\end{subfigure}
	\caption{The rotation of shear axes of geodesic null congruences for several values of the $\alpha$ parameter. The antinodes and nodes correspond to solid and dashed vertical lines, respectively. The initial conditions are set to have at the first antinode $z=\pi\lambda/2$: $arg(\sigma)=\pi$. The remaining parameters are $\lambda=1/10$, $\beta=3$. We set $t_0=0.383171$ which gives $z_0=0.226091$.}
\label{fig4}
\end{center}
\end{figure}

\begin{figure}[t!]
\begin{center}
\begin{subfigure}{.5\textwidth}
%\centering
\begin{flushleft}
\includegraphics[width=10cm,angle=0]{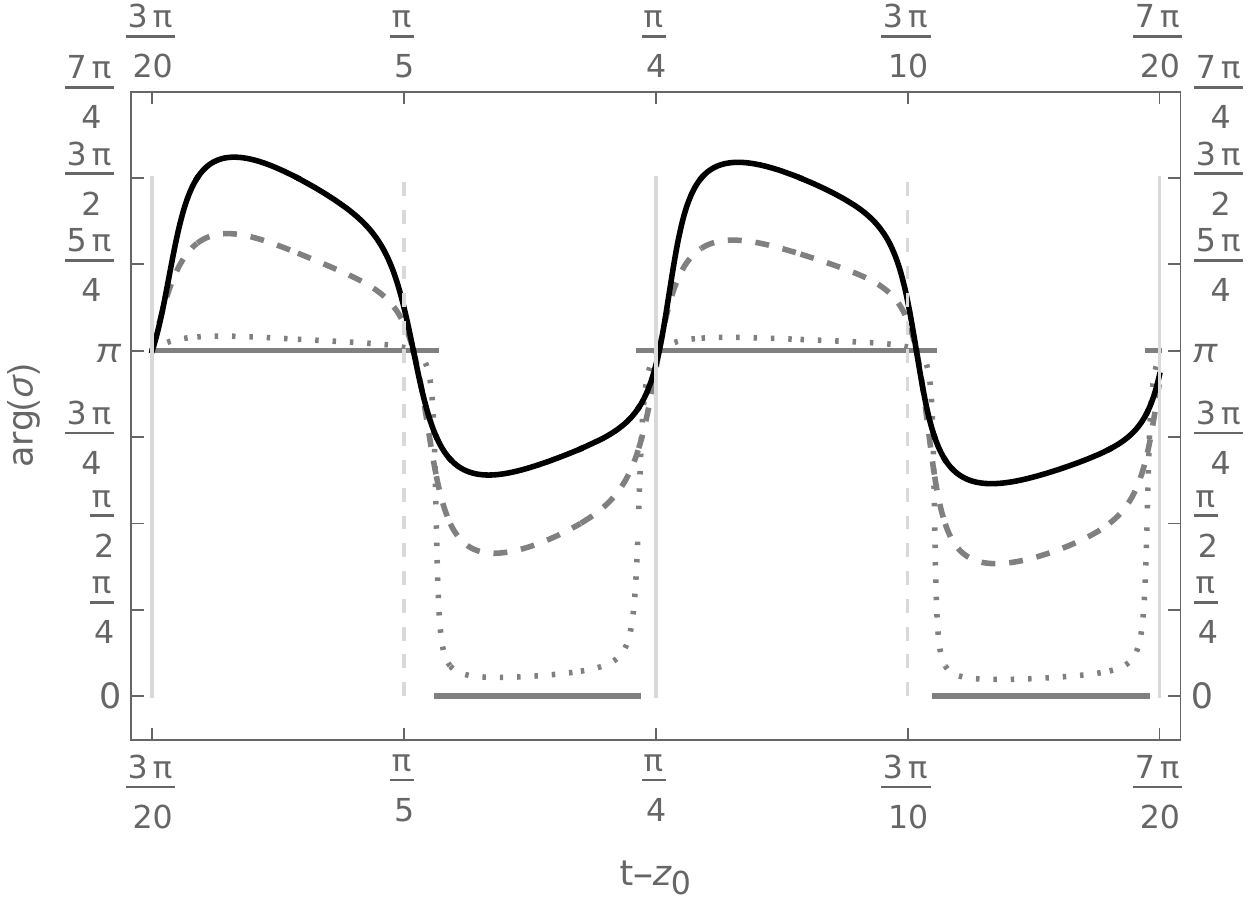}
\end{flushleft}
\end{subfigure}%
\begin{subfigure}{.5\textwidth}
\begin{flushright}
\includegraphics[width=3.3cm,angle=0]{fig4a.pdf}
\end{flushright}
\end{subfigure}
\caption{The rotation of shear axes of geodesic null congruences for several values of the $\alpha$ parameter. The antinodes and nodes correspond to solid and dashed vertical lines, respectively. The initial conditions are set to have at the second antinode $z=3\pi\lambda/2$: $arg(\sigma)=\pi$. The remaining parameters are $\lambda=1/10$, $\beta=3$. We set $t_0=0.383171$ which gives $z_0=-0.0880683$.
}
\label{fig5}
\end{center}
\end{figure}

Let us consider a light beam sent by one of the stationary observers at antinodes [$z=\pi\lambda(1/2+k)$, $k\in \mathbb{Z}$] along the $z$ direction at some initial moment $t_0$. The beam is detected by other stationary observers on subsequent antinodes. The equation for the metric function $p$ depends on the sign of $\sin(z/\lambda)$ which at antinodes equals $\pm 1$. Therefore, there are two types of antinodes in our model. They corresponds to odd and even values of $k$ and must be investigated separately. Without loss of generality we will consider below only $k=0$ and $k=1$. It follows from the form of the metric \ref{metric} that along such null rays $t=z+z_0$. The unknown constant $z_0$ can be chosen to have $arg(\sigma)=\pi$ at $t_0$ and $z=\pi\lambda/2$ ($k=0$) or $z=3\pi\lambda/2$ ($k=1$). These conditions imply $J_1(z_0+\pi\lambda/2)=0$ for $k=0$ and $J_1(z_0+3\pi\lambda/2)=0$ for $k=1$ (for $t_0=\pi\lambda/2+z_0$ and $t_0=3\pi\lambda/2+z_0$, respectively). The rotation of the shear axis of null congruences [given by $arg(\sigma)$] are presented in both cases in Figs.\ \ref{fig4} and \ref{fig5}. They are plotted for four different values of the parameter $\alpha$, which defines the polarization of gravitational waves. The value $\alpha=0$ corresponds to + polarized gravitational waves for which the shear axis of geodesic null congruence does not rotate. The square waveform ($\alpha=0$) oscillates between $0$ and $\pi$ in both figures, which implies that the congruence is stretched or contracted along the $x$ axis. The real part of the shear $\sigma$ does not change sign exactly at nodes and antinodes as is clearly visible in Figs.\ \ref{fig4} and \ref{fig5} for $\alpha=0$. The cases $k=0$ and $k=1$ correspond to a different direction of rotation of the shear axes after initial moment $t_0$. 

\begin{figure}[t!]
\begin{center}
%\centering
\includegraphics[width=10cm,angle=0]{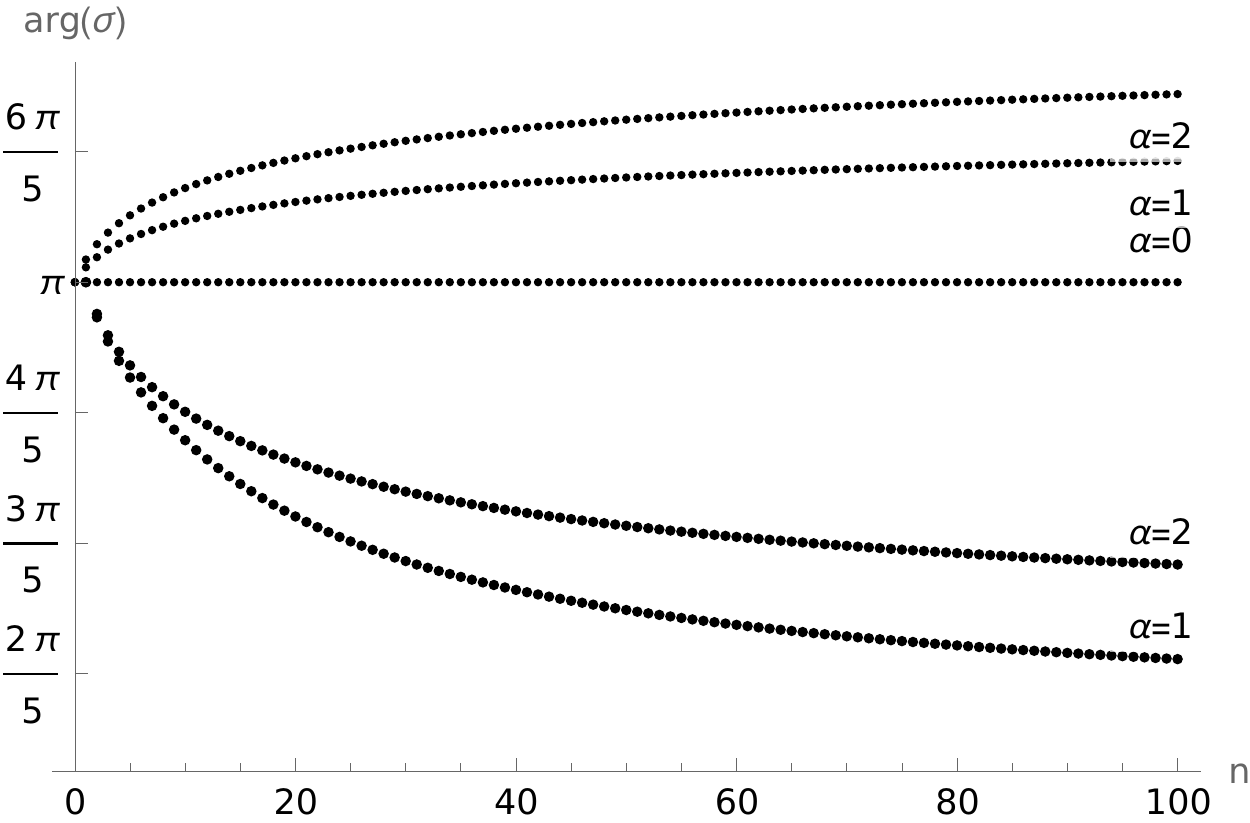}
	\caption{The cumulative effect of rotation of shear axes as measured at subsequent antinodes. The points above and at $\pi$ correspond to $k=0$. The points below and at $\pi$ correspond to $k=1$. The parameters are $\lambda=1/10$, $\beta=3$.}
\label{fig6}
\end{center}
\end{figure}

The oscillations of the shear axes visible in Figs.\ \ref{fig4} and \ref{fig5} suggest that the standing gravitational waves studied in this article are polarization standing waves. A more detailed analysis of these figures reveals that for $\alpha\neq 0$ the shear axes do not return to initial position [$arg(\sigma)=\pi$] at the subsequent nodes (Fig.\ \ref{fig6}).\footnote{It seems that the expansion of our cross-polarized standing wave spacetime is necessary to ``desynchronize'' polarization axes of gravitational waves and the shear axes of a null congruence.} This cumulative effect of the polarization of gravitational waves on shear axes of geodesic null congruences is observable by stationary observers at antinodes. Therefore, the polarization of nonlinear standing gravitational waves directly alters observables in spacetime.

\section{Interpretation of the $\alpha$ parameter}

The role of the $\alpha$ parameter follows from the field equations \cite{PiranSafier:1985}. A standing wave may be seen as a ``superposition'' of two identical gravitational waves moving in opposite directions. The amplitudes of these waves may be split between two polarizations. The oscillations of the metric function $\omega$ correspond to the rotation of polarization between $+$ and $\times$ modes. This is a gravitational analog of the Faraday rotation \cite{PiranSafier:1985}. Since $\omega=0$ at antinodes, then the trajectories of test particles at antinodes depend only indirectly on $\alpha$. One may try to understand this dependency in terms of the high-frequency limit of standing waves (e.g., Ref.\ \cite{ers}), where the energy of the gravitational waves alters global expansion of spacetime, but such attempts lead to counterintuitive results. 
\begin{figure}[t!]
\begin{center}
\includegraphics[width=10cm,angle=0]{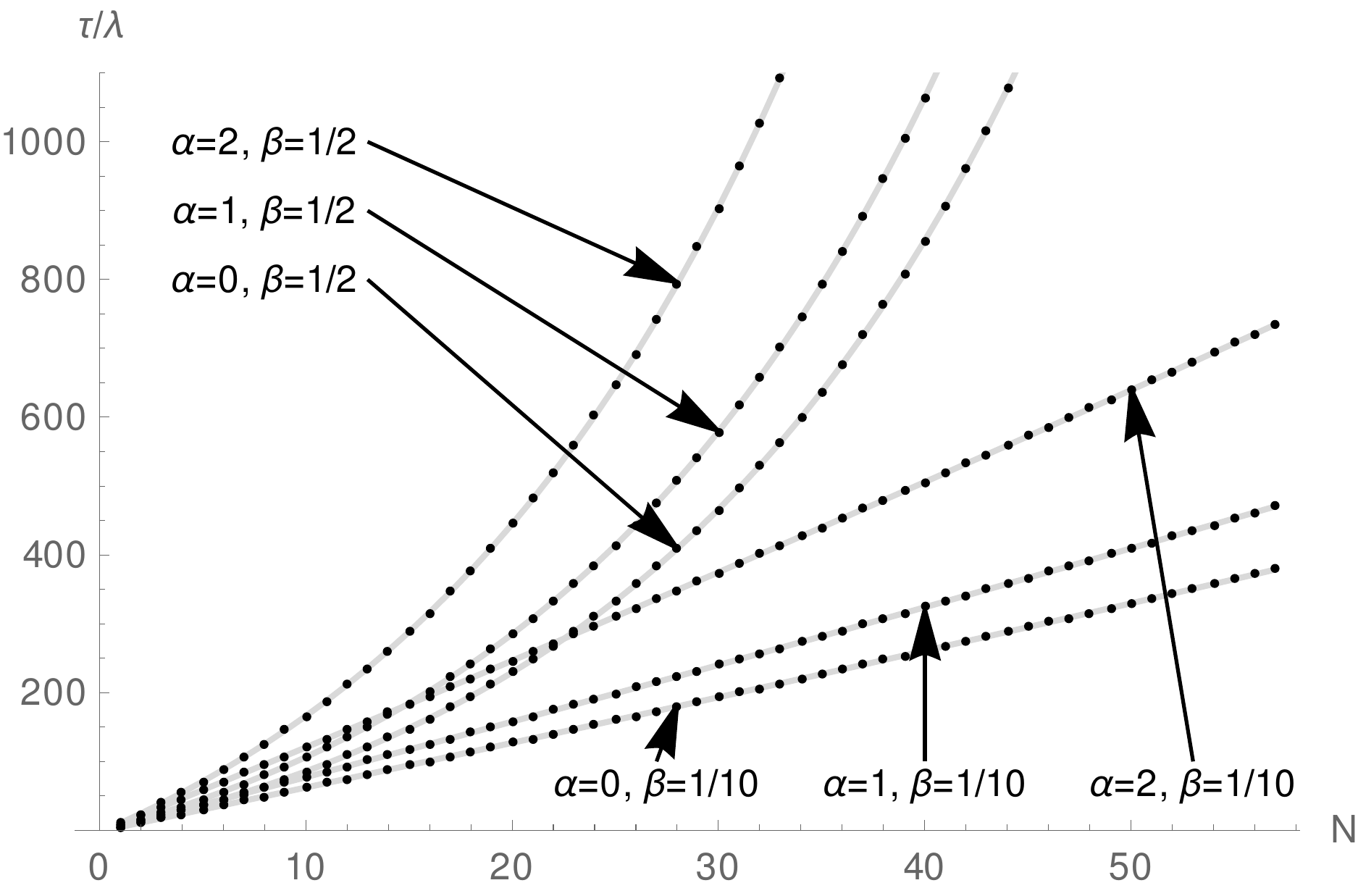}
\caption{The photon's flight time between subsequent antinodes (numbered by $N$) measured in the proper time of stationary observers at antinodes. The photon flies along the $\partial_z$ direction. The remaining parameters are $t_0=1$, $\lambda=1/10$.}
\label{fig1}
\end{center}
\end{figure}

In order to evaluate the effect of the $\alpha$ parameter on the global expansion explicitly, we calculated numerically a flight time of a photon between stationary observers at antinodes in terms of the proper time of these observers. We note that for $\beta=0$ the metric \eqref{metric} describes the three-torus identified Minkowski spacetime in an expanding coordinate system. Thus, any expansion for $\beta=0$ is of an artificial nature and depends on the choice of coordinates. Standing waves appear already for small $\beta$ and a position of stationary observers at antinodes is distinguished by the geometry of spacetime. We show in Fig.\ \ref{fig1} that in this setting the time flight of photons between subsequent observers (measured in observers' proper time) grows initially almost linearly with the slope controlled by $\alpha$ parameter. Therefore, $\alpha$ describes not only polarization of waves, but also encodes initial conditions that define how fast the distance between antinodes grows. Whether this is merely a coincidence (or an artifact of the model studied) needs further investigation. For larger $\beta$ an exponential expansion of spacetime becomes evident (Fig.\ \ref{fig1}), but the role of the $\alpha$ parameter remains unchanged: larger $\alpha$ implies faster expansion.

\section{The Poynting vector}

\begin{figure}[t!]
\begin{center}
\includegraphics[width=14.5cm,angle=0]{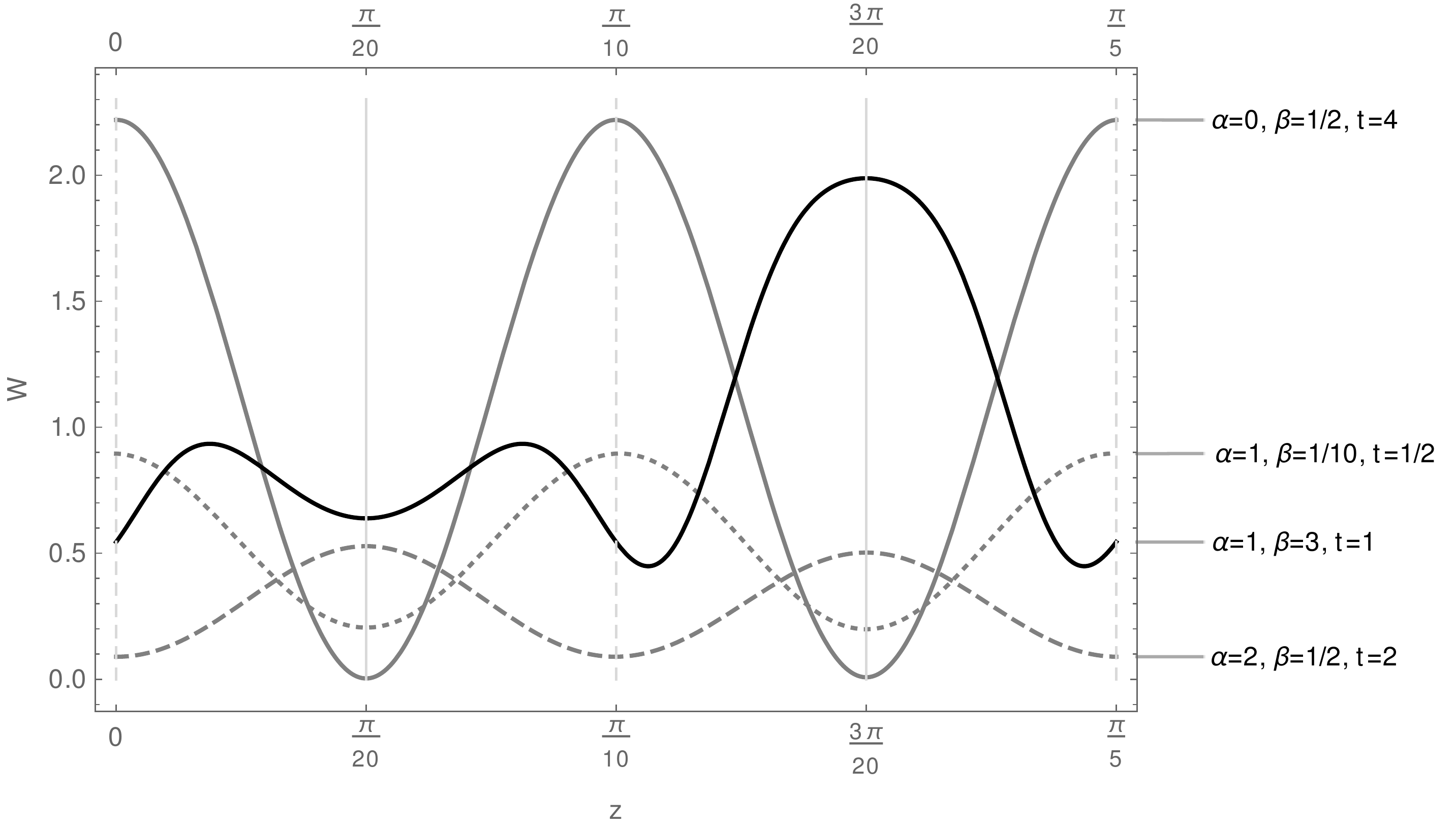}
\caption{The superenergy density for $\lambda=1/10$ and several sets of remaining parameters. The amplitude of the solid black function was multiplied by a factor of $300$. The antinodes and nodes are indicated with solid and dashed vertical lines, respectively.}
\label{fig2}
\end{center}
\end{figure}

\begin{figure}[t!]
\begin{center}
\begin{subfigure}{.5\textwidth}
%\centering
\begin{flushleft}
\includegraphics[width=10cm,angle=0]{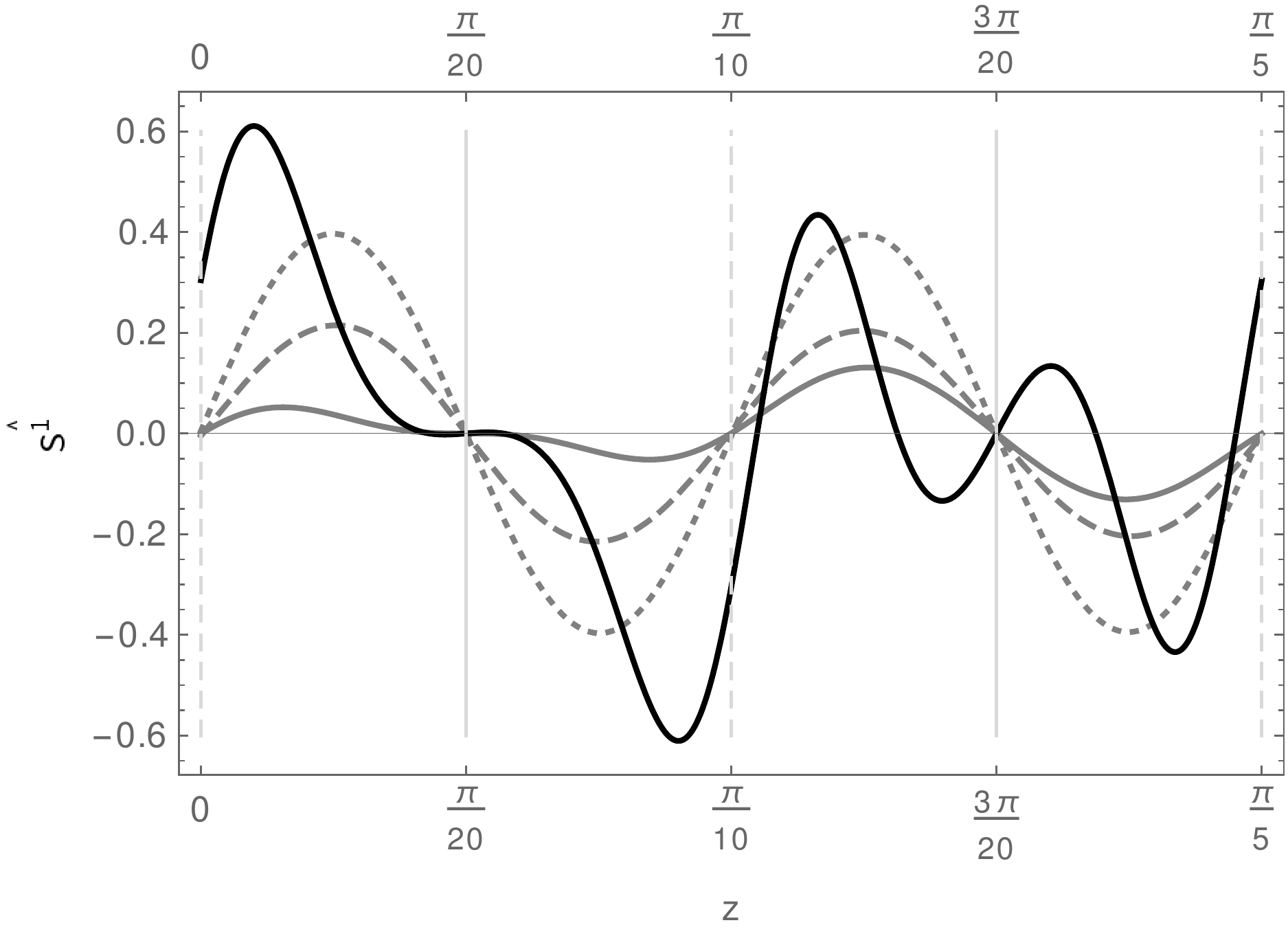}
\end{flushleft}
\end{subfigure}%
\begin{subfigure}{.5\textwidth}
\begin{flushright}
\includegraphics[width=3.3cm,angle=0]{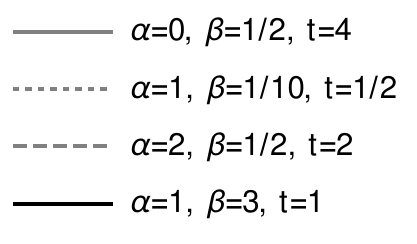}
\end{flushright}
\end{subfigure}
\caption{The $S^{\hat 1}$ component of the super-Poynting vector for $\lambda=1/10$ and several sets of remaining parameters. It shows the superenergy transfer in the $\partial_z$ direction. The amplitude of the solid black function was multiplied by a factor of $300$. The antinodes and nodes are indicated with solid and dashed vertical lines, respectively.}
\label{fig3}
\end{center}
\end{figure}

Stephani suggested \cite{Stephani:2003} that the gravitational analog of the Poynting vector may play an important role in the understanding of standing gravitational waves. The gravity is a nonlocal concept, so in general relativity the density of gravitational energy cannot be defined. Similarly, the flow gravitational energy can be described only in an approximate or asymptotic form. Therefore, one should refrain from a direct physical interpretation of the gravitational analogs of the Poynting vector. Nevertheless, we apply this kind of concept (a super-Poynting vector and a superenergy) as mathematical tools in a context of standing gravitational waves. Stephani derived the gravitational analog of the Poynting vector from the Lagrangian that leads to the Ernst equations. We follow another approach here. We construct the analog of the Poynting vector with the help of the Bel-Robinson tensor $T_{\alpha\beta\gamma\delta}$ \cite{Bel:1958}. The Bel-Robinson tensor is given by
\begin{equation}\label{BR}
T_{\alpha\beta\gamma\delta}= C_{\alpha\mu\gamma}^{\quad\nu}C_{\delta\nu\beta}^{\quad\mu} + \star C_{\alpha\mu\gamma}^{\quad\nu}\star C_{\delta\nu\beta}^{\quad\mu}\;.
\end{equation}	
$C$ is the Weyl tensor and $\star$ denotes the Hodge dual, namely, $$\star C_{\alpha\beta\gamma\delta}=\frac{1}{2}\eta_{\alpha\beta\mu\nu}C^{\mu\nu}_{\;\;\;\;\gamma\delta}\;,$$ where $\eta$ is the canonical volume form. The supermomentum $P$ is defined relative to an observer with four-velocity $u$: $P^\alpha=-T^{\alpha}_{\;\;\beta\gamma\delta}u^\beta u^\gamma u^\delta$. It might be decomposed into the superenergy density $W$ (which is not an energy density of the gravitational field) and the super-Poynting vector $S$: $P^\alpha=W u^\alpha+S^\alpha$. The superenergy density and the super-Poynting vector can be written in terms of the Bel-Robinson tensor,
\begin{equation}
\begin{split}
W=&T_{\alpha\beta\gamma\delta}u^\alpha u^\beta u^\gamma u^\delta\geq 0\;,\\
S^\alpha=&-T^\mu_{\;\;\beta\gamma\delta}(\delta^{\alpha}_{\;\mu}+u^\alpha u_\nu)u^\beta u^\gamma u^\delta\;.
\end{split}
\end{equation}
In the orthonormal frame, relative to the observer with the four-velocity $e_{\hat 0}$, these formulas take a form ($S^{\hat 0}=0$),
\begin{equation}
\begin{split}
W=&T_{\hat 0\hat 0\hat 0\hat 0}\;,\\
S^{\hat i}=&-T_{\hat i\hat 0\hat 0\hat 0}\;.
\end{split}
\end{equation}
The expressions for the superenergy $W$ and the super-Poynting vector for solutions studied in this paper are too long to be usefully cited here. The analysis of these formulas with the help of the computer algebra system {\sc Mathematica} reveals
\begin{itemize}
\item $W$ oscillates periodically with extrema at antinodes,
\item $S^{\hat 1}$ oscillates periodically with zeros at antinodes (no superenergy transfer at antinodes),
\item $S^{\hat 1}$ averages to zero over hypersurfaces $t=const$,
\item $S^{\hat 2}=S^{\hat 3}=0$.
\end{itemize}
We remind the reader here that beyond antinodes observers with the four-velocity $e_{\hat 0}$ do not move on geodesics. The superenergy and the $z$ component of the super-Poynting vector are presented in Figs.\ \ref{fig2} and \ref{fig3} for a particular set of parameters.


\section{Summary}

We constructed, starting from the Halilsoy cylindrical spacetime \cite{Halilsoy:1988}, the cosmological model of the Gowdy form with standing gravitational waves. This solution generalizes the spacetime studied in \cite{SzybkaNaqvi:2021} (with + polarization) to cross-polarized waves (denoted with $\times$). Similar to the + case, in the $\times$ polarized model there exist stationary observers at antinodes. The trajectories of neighboring massive test particles for these observers differ trivially from the trajectories in the + polarized model. The Tissot diagrams do not reveal standard effects of cross polarization on massive particles (although the $\times$ polarized model expands differently than the + polarized model). In order to show these effects, we studied the massless test particles.

We showed that the deviation of the model from the + polarization alters the shear axes of geodesic null congruences. The cross polarization of the standing gravitational waves can be directly detected by stationary observers at antinodes via the measurements of the rotation of the shear axes of geodesic null congruences aligned with the longitudinal direction of the standing gravitational waves. This constitutes the main result of our paper.

The deviation from the + polarized models alters also the expansion rate of the spacetime. In order to evaluate this effect, we studied the time of flight of photons between antinodes measured in the proper time of stationary observers at antinodes. We showed that the expansion rate of the model depends on a parameter that describes the deviation of the model from the + polarized model and controls the polarization. Finally, we used the Bel-Robinson tensor to calculate the superenergy density and the gravitational analog of the Poynting vector. We showed that both quantities have expected properties and are useful tools in the studies of the standing gravitational waves (however, we remind the reader that for fundamental reasons they cannot be identified with the density and flux of gravitational radiation). 

%\vspace{0.2cm}

%\noindent{\sc Acknowledgments}

%\clearpage
\appendix
\setcounter{equation}{0}
\renewcommand{\theequation}{A\arabic{equation}}

\section{From Halilsoy solution to cross-polarized Gowdy standing waves}
\label{A1}

Halilsoy showed \cite{Halilsoy:1988} that the Einstein-Rosen waves \cite{EinsteinRosen37,Rosen54} may be extended to include the second polarization. The exact solution to Einstein equations has a form
\begin{equation}\label{metricH}
g=e^{2(\gamma-\psi)}\left(-dt^2+d\rho^2\right)+\rho^2e^{-2\psi}d\varphi^2+e^{2\psi}(dz+\omega d\varphi)^2\;,
\end{equation}
where $\rho>0$, $-\infty<t,z<\infty$, and $0\leq\varphi<2\pi$. The metric functions $\psi$, $\gamma$, and $\omega$ depend on $t$ and $\rho$ only and are given by
\begin{equation}\label{solH}
\begin{split}
e^{-2\psi}=&e^{A J_0 \cos{\sigma t}}\sinh^2\frac{\alpha}{2}+e^{-AJ_0 \cos\sigma t}\cosh^2\frac{\alpha}{2}\;,\\
\omega=&-(A\sinh\alpha)\rho J_1\sin\sigma t\;,\\
\gamma=&\frac{1}{8}A^2\left(\sigma^2\rho^2(J_0^2+J_1^2)-2\sigma\rho J_0 J_1\cos^2\sigma t\right)\;,
\end{split}
\end{equation}
where $\sigma> 0$ is a constant, and $J_0=J_0(\sigma\rho)$ and $J_1=J_1(\sigma\rho)$ are the Bessel functions of the first kind and orders $0$ and $1$, respectively.

The Einstein equations for the metric \eqref{metricH} are form invariant under the complex substitution
$$t\mapsto i z,\;,\qquad\rho\mapsto i t\;,\qquad\varphi\mapsto i y\;,\qquad z\mapsto z\;,\qquad\omega\mapsto -i\omega$$ 
which brings the metric \eqref{metricH} into the form \eqref{metric}, where the remaining metric functions are related by $f=2(\gamma-\psi)$ and $p=-\ln t+2\psi$. Therefore, any cylindrical solution has its cosmological counterpart. The Halilsoy solution \eqref{solH} corresponds to \eqref{sol} with additional trivial redefinitions: $\sigma\mapsto 1/\lambda$ followed by $z\mapsto z+\lambda \pi/2$ and $A\mapsto 2\beta\sqrt{\lambda}$.

\setcounter{equation}{0}
\renewcommand{\theequation}{B\arabic{equation}}

\section{Geodesic equation}\label{A2}

The geodesic equation in coordinates $x^\alpha=(t,z,x,y)$ for the metric \eqref{metric} has a form
\begin{equation}
\begin{split}\label{eq:geo}
\ddot t &+\frac{1}{2}\left\{f_{,t} \dot{t}^2+2f_{,z}\dot t \dot z + f_{,t} \dot{z}^2+e^{-f}\left[e^{p}\dot{x}^2(1+p_{,t}t)+e^{-p}\dot{y}^2(1-p_{,t}t)\right]\right\}\\
&+e^{p-f}\left(t\omega_{,t}\dot{x}\dot{y}+\omega \dot{y}[(1+p_{,t}t)+t\omega_{,t}]+\frac{1}{2}\omega^2\dot z(1+p_{,t}t)\right)=0\;,\\
\ddot z &+\frac{1}{2}\left[f_{,z} \dot{z}^2+2f_{,t} \dot t \dot z + f_{,z} \dot{t}^2+t p_{,z}e^{-f}(-e^{p}\dot{x}^2+e^{-p}\dot{y}^2)\right]\\
&-e^{p-f}\dot{y}\left(t\dot{x}\omega_{,z}+\omega\dot{x}p_{,z}+\dot{y}\omega_{,z}+\frac{1}{2}\omega^2t \dot{y} p_{,z}\right)=0\;,\\
\ddot x &+\dot{x}(\dot{t}/t+p_{,t}\dot{t}+p_{,z}\dot{z})+\dot{y}(\dot{t}\omega_{,t}+\dot{z}\omega_{,z})+\omega\left[2\dot{y}(\dot{t}p_{,t}+\dot{z}p_{,z})-e^{2p}\dot{x}(\dot{t}\omega_{,t}+\dot{z}\omega_{,z})\right]\\
&-e^{2p}\omega^2\dot{y}(\dot{t}\omega_{,t}+\dot{z}\omega_{,z})=0\;,\\
\ddot y &+\dot{y}(\dot{t}/t-p_{,t}\dot{t}-p_{,z}\dot{z})+e^{2p}(\dot{x}+\omega \dot{y})(\dot{t}\omega_{,t}+\dot{z}\omega_{,z})=0\;,
\end{split}
\end{equation}
where a dot denotes differentiation in the proper time $\tau$ or the affine parameter for timelike or null geodesics, respectively. The normalization of the four-velocity/wave vector gives rise to the first integral of the form
\begin{equation}\nonumber
-\epsilon=e^f(-\dot{t}^2+\dot{z}^2)+t(e^p(\dot{x}+\omega\dot{y})^2+e^{-p}\dot{y}^2)\;,
\end{equation}
where the constant $\epsilon$ is equal to $1$ or $0$ for timelike or null geodesics, respectively. The Killing fields $\partial_x$ and $\partial_y$ give two more quantities $c_x$ and $c_y$ that are conserved along geodesics
\begin{equation}
\begin{split}\label{fi}
(\dot x+\omega\dot y) e^p t&=c_x\;,\\
\dot y e^{-p} t+ (\dot x+\omega\dot y)\omega e^p t&=c_y\;.
\end{split}
\end{equation}

\setcounter{equation}{0}
\renewcommand{\theequation}{C\arabic{equation}}

\section{Observers at antinodes}\label{A3}

In this appendix, we show that the observers at antinodes are stationary. The geodesic equation is presented in Appendix \ref{A2}. For stationary solutions $\dot z=\dot x=\dot y=0$ it takes the same form as for $\alpha=0$, thus this part of the analysis mimics calculations in \cite{SzybkaNaqvi:2021}. For readers' convenience, we repeat it below. We have
\begin{eqnarray}
        0&=&\ddot{t}+\frac{1}{2}f_{,t}\dot{t}^2\;,\label{fc1}\\
        0&=&f_{,z}\;,\nonumber\\
        \dot{t}&=&e^{-\frac{1}{2}f}\;,\label{fc3}
\end{eqnarray}
where \eqref{fc3} is the first integral of \eqref{fc1}. The condition $f_{,z}=0$ corresponds to antinodes. Using \eqref{sol} we have $f_{,z}=(\dots)\cos{(z/\lambda)}$, thus $z=\lambda\pi(1/2+k)$, where $k\in\mathbb{Z}$, implies $f_{,z}=0$. Therefore, the curve $\gamma_k$,
\begin{equation}\label{fge}
x^\mu=[t(\tau),\lambda\pi(1/2+ k),x_0,y_0]\;,
\end{equation}
with $k\in\mathbb{Z}$ and $t(\tau)$ determined by \eqref{fc3} is a future-directed timelike geodesic and a stationary solution to the geodesic equation. 

\setcounter{equation}{0}
\renewcommand{\theequation}{D\arabic{equation}}

\section{Orthonormal cobasis and its external derivative}\label{A4}

\begin{equation}
\begin{split}
\theta^{\hat{0}}=&e^{f/2}\,dt\;,\\
\theta^{\hat{1}}=&e^{f/2}\,dz\;,\\
\theta^{\hat{2}}=&\sqrt{t}\,e^{p/2}\,\left(dx+\omega\,dy \right)\;,\\
\theta^{\hat{3}}=&\sqrt{t}\,e^{-p/2}\,dy\;,\\
d\theta^{\hat{0}}=&-\frac{1}{2}e^{-f/2}\,f_{,z}\,\theta^{\hat{0}}\wedge \theta^{\hat{1}}\;,\\
d\theta^{\hat{1}}=&\frac{1}{2}e^{-f/2}\,f_{,t}\,\theta^{\hat{0}}\wedge \theta^{\hat{1}}\;,\\
d\theta^{\hat{2}}=&\frac{1}{2}e^{-f/2}\left(\left(\frac{1}{t}+p_{,t} \right)\,\theta^{\hat{0}}\wedge \theta^{\hat{2}}+
p_{,z}\,\theta^{\hat{1}}\wedge \theta^{\hat{2}}
+2e^{p}\left(\omega_{,t}\,\theta^{\hat{0}}\wedge \theta^{\hat{3}}+\omega_{,z}\,\theta^{\hat{1}}\wedge \theta^{\hat{3}} \right)\right)\;,\\
d\theta^{\hat{3}}=&\frac{1}{2}e^{-f/2}\left(\left(\frac{1}{t}-p_{,t} \right)\,\theta^{\hat{0}}\wedge \theta^{\hat{3}}
-p_{,z}\,\theta^{\hat{1}}\wedge \theta^{\hat{3}} \right)\;.
\end{split}
\end{equation}

\setcounter{equation}{0}
\renewcommand{\theequation}{E\arabic{equation}}

\section{Connection 1-forms}\label{A5}

The nonzero connection 1-forms in the orthonormal frame are as follows:

\begin{equation}
\begin{split}
\omega^{\hat{0}}\,{}_{\hat{1}}=&\frac{1}{2} e^{-f/2}\left(f_{,z}\,\theta^{\hat{0}}+f_{,t}\,\theta^{\hat{1}} \right)\;,\\
\omega^{\hat{0}}\,{}_{\hat{2}}=&\frac{1}{2}e^{-f/2}\left(\left(\frac{1}{t}+p_{,t} \right)\,\theta^{\hat{2}}+e^p\,\omega_{,t}\theta^{\hat{3}}\right)\;,\\
\omega^{\hat{0}}\,{}_{\hat{3}}=&\frac{1}{2}e^{-f/2}\left(\left(\frac{1}{t}-p_{,t} \right)\,\theta^{\hat{3}}+e^p\,\omega_{,t}\theta^{\hat{2}}\right)\;,\\
\omega^{\hat{1}}\,{}_{\hat{2}}=&-\frac{1}{2}e^{-f/2}\left(e^p\,\omega_{,z}\theta^{\hat{3}}+p_{,z}\,\theta^{\hat{2}}\right)\;,\\
\omega^{\hat{1}}\,{}_{\hat{3}}=&\frac{1}{2}e^{-f/2}\left(p_{,z}\,\theta^{\hat{3}}-e^p\,\omega_{,z}\theta^{\hat{2}}\right)\;,\\
\omega^{\hat{2}}\,{}_{\hat{3}}=&-\frac{1}{2}e^{-f/2}\,e^{p}\,\left(\omega_{,t}\,\theta^{\hat{0}}+\omega_{,z}\,\theta^{\hat{1}} \right)\;.
\end{split}
\end{equation}

\setcounter{equation}{0}
\renewcommand{\theequation}{F\arabic{equation}}

\section{Curvature 2-forms}\label{A6}

The nonzero curvature 2-forms in the orthonormal frame are as follows:

\begin{equation*}
\begin{split}
\Omega^{\hat{1}}\,{}_{\hat{0}}=&\frac{1}{2}e^{-f} \left[(f_{,tt}-f_{,zz})\theta^{\hat{0}}\wedge \theta^{\hat{1}}+
e^p\,(p_{,t}\omega_{,z}-p_{,z}\omega_{,t})\theta^{\hat{2}}\wedge \theta^{\hat{3}}\right]\;,\\
\Omega^{\hat{2}}\,{}_{\hat{0}}=&\frac{1}{4}e^{-f} \left[\left((p_{,t}+\frac{1}{t})(p_{,t}-f_{,t}+\frac{1}{t})+2(p_{,tt}-\frac{1}{t^2})-p_{,z}f_{,z}
-e^{2p}\omega_{,t}^2 \right)\theta^{\hat{0}}\wedge \theta^{\hat{2}}-\right. \\
&\left.-e^p\,\left(\omega_{,t}(f_{,t}-4p_{,t}-\frac{2}{t})-2\omega_{,tt}+\omega_{,z}f_{,z} \right)\theta^{\hat{0}}\wedge \theta^{\hat{3}}+\right. \\ 
&\left.+\left((p_{,z}-f_{,z})(p_{,t}+\frac{1}{t})+2p_{,tz}-p_{,z}f_{,t}-e^{2p}\omega_{,t}\omega_{,z} \right)\theta^{\hat{1}}\wedge \theta^{\hat{2}}-\right. \\
&\left.-e^p\,\left(\omega_{,t}(f_{,z}-p_{,z})-2p_{,t}\omega_{,z}-2\omega_{,tz}+\omega_{,z}(f_{,t}-p_{,t}+\frac{1}{t}) \right)\theta^{\hat{1}}\wedge \theta^{\hat{3}} \right]\;,
\end{split}
\end{equation*}
\begin{equation*}
\begin{split}
\Omega^{\hat{3}}\,{}_{\hat{0}}=&\frac{1}{4}e^{-f} \left[-e^p\,\left(\omega_{,t}(f_{,t}-4p_{,t}-\frac{2}{t})-2\omega_{,tt}+f_{,z}\omega_{,z} \right)\theta^{\hat{0}}\wedge \theta^{\hat{2}}+ \right.\\
&+\left.\left((f_{,t}+p_{,t}-\frac{1}{t})(p_{,t}-\frac{1}{t})-2(p_{,tt}+\frac{1}{t^2})+p_{,z}f_{,z}+3e^{2p}\omega_{,t}^2 \right)\theta^{\hat{0}}\wedge \theta^{\hat{3}}-\right. \\ 
&\left.-e^p\,\left(\omega_{,t}(f_{,z}-3p_{,z})-2\omega_{,tz}+\omega_{,z}(f_{,t}-p_{,t}-\frac{1}{t}) \right)\theta^{\hat{1}}\wedge \theta^{\hat{2}}+\right. \\
&\left.+\left((f_{,z}+p_{,z})(p_{,t}-\frac{1}{t})-2p_{,tz}+p_{,z}f_{,t}+3e^{2p}\omega_{,t}\omega_{,z} \right)\theta^{\hat{1}}\wedge \theta^{\hat{3}} \right]\;.
\end{split}
\end{equation*}

\setcounter{equation}{0}
\renewcommand{\theequation}{G\arabic{equation}}

\section{Riemann tensor}\label{A7}

Independent nonzero components of the Riemann tensor in the orthonormal frame are as follows:
\begin{equation}
\begin{split}\label{Riemann}
R^{\hat 1}_{\;\hat 0\hat 1\hat 0}&=\frac{1}{2}e^{-f}(f_{,zz}-f_{,tt})\;,\\
	R^{\hat 2}_{\;\hat 0\hat 2\hat 0}&=-\frac{1}{4}e^{-f}\left[(p_{,t}+\frac{1}{t})^2-f_{,t}(p_{,t}+\frac{1}{t})+2(p_{,tt}-\frac{1}{t^2})-f_{,z}p_{,z}-e^{2p}\omega_{,t}^2\right]\;,\\
	R^{\hat 3}_{\;\hat 0\hat 3\hat 0}&=-\frac{1}{4}e^{-f}\left[(p_{,t}-\frac{1}{t})^2+f_{,t}(p_{,t}-\frac{1}{t})-2(p_{,tt}+\frac{1}{t^2})+f_{,z}p_{,z}+3e^{2p}\omega_{,t}^2\right]\;,\\
	R^{\hat 2}_{\;\hat 0\hat 3\hat 0}&=\frac{1}{4}e^{-f+p}\left[(-2/t+f_{,t}-4p_{,t})\omega_{,t}-2\omega_{,tt}+f_{,z}\omega_{,z}\right]\;.
\end{split}
\end{equation}

\bibliographystyle{apsrev}
\setcitestyle{authortitle}
\bibliography{report}

\begin{thebibliography}{10}

\bibitem{Hermann:2004}
H.~Bondi,
\newblock Gravitational waves in general relativity {XVI}. {S}tanding waves,		
\newblock \href{https://doi.org/10.1098/rspa.2003.1176}{{ Proc.\ R. Soc.\ A} 460, 463 (2004)}.

\bibitem{Stephani:2003}
H.~Stephani,
\newblock Some remarks on standing gravitational waves,
\newblock \href{https://doi.org/10.1023/A:1022330218708}{{ Gen. Relativ.\ Gravit.} 35, 467 (2003)}.

\bibitem{Szybka:2019}
S.~J. Szybka and A.~Cieślik,
\newblock {Standing waves in general relativity},
		\newblock \href{https://doi.org/10.1103/PhysRevD.100.064025}{{ Phys.\ Rev.\ D} 100, 064025 (2019)}.

\bibitem{SzybkaNaqvi:2021}
S.~J. Szybka and S.~U. Naqvi,
\newblock Freely falling bodies in a standing-wave spacetime,
\newblock \href{https://doi.org/10.1103/PhysRevD.103.024011}{{ Phys.\ Rev.\ D} 103, 024011 (2021)}.

\bibitem{Halilsoy:1988}
M.~Halilsoy,
\newblock {Cross-polarized cylindrical gravitational waves of Einstein and
  Rosen,}
\newblock \href{https://doi.org/10.1007/BF02725615}{{ Nuovo Cimento B Ser.} 102, 563 (1988)}.

\bibitem{gpw}
X. Fang, K. F. MacDonald, E. Plum, and N. I. Zheludev,
\newblock{Coherent control of light-matter interactions in polarization
standing waves,}
\newblock \href{https://doi.org/10.1038/srep31141}{{ Sci. Rep.} 6, 31141 (2016)}.

\bibitem{tidal}
R. Goswami and G. F. R. Ellis,
\newblock{Tidal forces are gravitational waves,}
\newblock \href{https://doi.org/10.1088/1361-6382/abdaf3}{{ Class.\ Quantum Grav.} 38, 085023 (2021)}.

\bibitem{PiranSafier:1985}
T. Piran and P. N. Safier,
\newblock A gravitational analogue of {F}araday rotation,
\newblock \href{https://doi.org/10.1038/318271a0}{{ Nature (London)} 318, 271 (1985)}.

\bibitem{ers}
S.~J. Szybka and M.~J. Wyr\ifmmode~\mbox{\k{e}}\else \k{e}\fi{}bowski,
\newblock Backreaction for {E}instein-{R}osen waves coupled to a massless
  scalar field,
\newblock \href{https://doi.org/10.1103/PhysRevD.94.024059}{{ Phys.\ Rev.\ D} 94, 024059 (2016)}.

\bibitem{Bel:1958}
L. Bel,
\newblock Sur la radiation gravitationnelle,
\newblock { C.R. Hebd. Seances Acad.\ Sci.} 247, 1094 (1958).

\bibitem{EinsteinRosen37}
A.~Einstein and N.~Rosen,
\newblock {On gravitational waves},
\newblock \href{https://doi.org/10.1016/S0016-0032(37)90583-0}{{ J. Franklin Inst.} 223, 43 (1937)}.

\bibitem{Rosen54}
N.~Rosen,
\newblock Some cylindrical gravitational waves,
\newblock {Bull.\ Res.\ Counc.\ Isr.} 3, 328 (1954).

\end{thebibliography}
\end{document}